\documentclass[12pt]{article}

\usepackage{sbc-template}

\usepackage{graphicx,url}
\usepackage[utf8]{inputenc}  

\usepackage{etoolbox}

\usepackage{multirow}
\usepackage{framed}
\usepackage{multicol}
\usepackage{booktabs}
\usepackage{graphicx,url}
\usepackage{courier}
\usepackage{xspace}
\usepackage{listings}
\usepackage{cite}
\usepackage{balance}
\usepackage{tablefootnote}
\usepackage{tabularx}
\usepackage{enumerate}
\usepackage{colortbl}
\usepackage{setspace}
\usepackage{booktabs}

\usepackage[table,x11names,dvipsnames,table]{xcolor}

\makeatletter
\patchcmd{\ttlh@hang}{\parindent\z@}{\parindent\z@\leavevmode}{}{}
\patchcmd{\ttlh@hang}{\noindent}{}{}{}
\makeatother

\definecolor{mygray}{gray}{0.4}

\newcommand{\aspas}[1]{{``#1''}}

\newcommand*{\titulo}[1]{%
\textcolor{black}{\textbf{#1}}%
}

\PassOptionsToPackage{hyphens}{url}
\usepackage{hyperref}
\hypersetup{%
  colorlinks=false,
  pdfborderstyle={/S/U/W 0}
}

\title{Monorepos: A Multivocal Literature Review}

\author{Gleison Brito\inst{1}, Ricardo Terra\inst{2}, Marco Tulio Valente\inst{1}}

\address{Federal University of Minas Gerais, Belo Horizonte, Brazil
 \nextinstitute
  Federal University of Lavras, Lavras, Brazil
 \email{\{gleison.brito,mtov\}@dcc.ufmg.br, terra@dcc.ufla.br}
}

\begin{document} 
\sloppy
\maketitle

\begin{abstract}
	Monorepos (Monolithic Repositories) are used by large companies, such as Google and Facebook, and by popular open-source projects, such as Babel and Ember. 
	This study provides an overview on the definition and characteristics of monorepos as well as on their benefits and challenges. 
	Thereupon, we conducted a multivocal literature review on mostly grey literature. 
	Our findings are fourfold. First, monorepos are single repositories that contains multiple projects, related or unrelated, sharing the same dependencies. Second, centralization and standardization are some key characteristics. Third, the main benefits include simplified dependencies, coordination of cross-project changes, and easy refactoring. Fourth, code health, codebase complexity, and tooling investments for both development and execution are considered the main challenges.
\end{abstract}

\section{Introduction}

Monorepos (Monolithic Repositories or Multi-Package Repositories) are commonly described as a single repository containing more than one project, in contrast to the single-repository-per-project model~\cite{tagMonoreposStackOverflow}. This model is adopted for several large software companies, including \emph{Facebook}, \emph{Google}, and \emph{Microsoft} \cite{facebookMonorepo, potvin2016google, microsoftMonorepo}. Some popular \emph{open-source} projects also adopt this model to manage their repositories (e.g., Babel and Ember).

The discussion about the adoption of monorepos is an emerging theme in the developer community. The theme is discussed in several forums and blogs, where the adoption of monorepos is either defended or rebutted. There are also much debate about the migration of multiple repositories to a single one, mainly motivated by the adoption of the monorepo model by large companies such as Google and Facebook. However, there is no consensus on the benefits and challenges of this repository model.

In view of such context,  this  paper  provides an overview on monorepos. Thereupon, we conduced a Multivocal Literature Review based on 21 grey literature and two academic papers in order to: (i) undestand the definitions of monorepo, (ii) identify characteristics of monorepos, (iii) investigate benefits of monorepos, and (iv) investigate challenges of monorepos.  

Regarding~(i), monorepos are usually defined as a single repository that contains multiple projects related or unrelated, but there are two kinds of monorepos: {``Monstrous'' monorepos}, which are commonly used by large companies, such as {Google} and {Facebook},  and {Projects monorepos}, which are used by medium-size open-source projects, such as React and {Babel}. 
Regarding~(ii), the projects into a Monorepo share the same managing tools and the same version of dependencies, besides all projects are visible to contributors.
Regarding~(iii), benefits include simplified dependencies, better managing of cross-project changes, easy refactoring, simplified organization, improve overall work culture, better coodination between developers, and better support of build tools to manage the repository. Regarding~(iv), challenges include difficulties with code heath, codebase complexity, tooling investiments, loss of version information, build, deploy and test tasks, and migration of many repositories to only one.
%

The remainder of this paper is organized as follows.  Section \ref{researchMetodology} describes our research metodology. Section \ref{results} reports the results of our multivocal literature review.  Section~\ref{threatsValidity} discusses the threats to validity of our study and Section \ref{conclusion} concludes.

\section{Research Metodology}
\label{researchMetodology}
In this section, we describe our research methodology. We also provide an overview 
of the systematic approach used to gather relevant literature.

\subsection{Literature Review}

After an initial search in the literature to learn more on the topic of monorepos,
we could not find a substantial body of academic research on the topic. We
therefore decided to conduct a Multivocal Literature Review (MLR), which is based on all accessible literature on a topic \cite{ogawa1991towards}. This includes---but
is not limited to---blogs, white papers, articles, and academic literature. 
By using
this wide spectrum of literature, the results will give a more broad view at the topic
since they include the voices and opinions of academics, practitioners, independent
researchers, development firms, and others who have experience on the topic~\cite{ogawa1991towards}.

Garousi et al.~\cite{garousi2016need} emphasize the importance of MLRs in the Software Engineering (SE) field
by stating that SE practitioners produces grey literature on a great scale,
but most are not published in academic vehicles. Therefore, they argue that
not including that literature in systematic reviews implies in researchers missing out important current state-of-the-art practice in SE.

\subsection{Research Questions}
\label{sec:rq}

We conducted this MLR to obtain an understanding of what a monorepo model is, what are the key characteristics of this model, and what are the benefits and challenges of adopting it.
In order to achieve the goal, we formulate four research questions:

\begin{enumerate}[\noindent\bf RQ \#1.]
\item 
How does the literature define monorepos?\\[-0.3cm]

\item 
What are the characteristics of monorepos?\\[-0.3cm]

\item 
What are the main expected benefits of adopting monorepos?\\[-0.3cm]

\item 
What are the main expected challenges of adopting monorepos?\\[-0.3cm]
\end{enumerate}


\subsection{Study Protocol}

This section describes the systematic protocol we followed to retrieve the literature used
in our study. 
We describe the databases, the search
strategy used to find related literature, 
the inclusion and exclusion criteria used to find the most relevant literature, 
and the process in which we catalogued the literature.\\

\noindent\textbf{Databases.} We relied on Google’s search engine to find relevant literature:

\begin{itemize}
	\item \textbf{Google Search}\footnote{http://www.google.com/} to locate grey literature (white
	papers, blogs, articles, etc.)\\[-0.3cm]
	\item \textbf{Google Scholar}\footnote{http://scholar.google.com/} to specifically locate academic literature.
\end{itemize}

We chose Google's search engines instead of more traditional search engines
(like Springer
Link, ACM Digital Library, IEEE Explore, etc.) because Monorepo is a very new
topic and limited academic research is available. We therefore knew before-hand that this literature review would rely mostly on grey literature.

\noindent\textbf{Search Terms.} The search string were built following the steps proposed by
Brereton et al.~\cite{brereton2007lessons}:

\begin{enumerate}
	\item Derive major terms from the research questions by identifying the main concepts.\\[-0.3cm]
	\item Identify alternative spellings and synonyms for major terms.\\[-0.3cm]
	\item Check the keywords in any relevant papers we already have.\\[-0.3cm]
	\item Use the boolean OR to add alternatives spellings and synonyms.\\[-0.3cm]
	\item Use the boolean AND to link the major terms.
\end{enumerate}

Since the search strings aims to find relevant literature related to the RQs, we defined
them as follows:\\[-0.2cm]


\begin{footnotesize}
\noindent\fbox{
\begin{minipage}{1\textwidth}
\texttt{("monorepo" \textbf{OR} "monolithic repository" \textbf{OR} "multi-package repository")\\
\hspace*{200pt}\textbf{AND} \\
("definition" \textbf{OR} "definitions" \textbf{OR}\\ 
\hspace*{6pt}"characteristic" \textbf{OR} "characteristics" \textbf{OR}\\
\hspace*{6pt}"benefit" \textbf{OR} "benefits" \textbf{OR} "challenge" \textbf{OR} "challenges")
}
\end{minipage}}
\end{footnotesize}

\noindent\textbf{Study Selection.} After retrieving the results of the initial search, we excluded irrelevant articles using the following inclusion and exclusion criteria:

\begin{itemize}
	\item Inclusion criteria:\\[-0.2cm]
		\begin{itemize}
			\item Literature that explicitly discuss monorepos;\\[-0.3cm]
			\item Literature that explicitly discuss the
			challenges and benefits of monorepos;\\[-0.3cm]
			\item Literature that discuss the definition of monorepos;\\[-0.3cm]
			\item Literature published after 2014; and\\[-0.3cm]
			\item Literature that appears in the five first pages on Google's search.\\[-0.2cm]
		\end{itemize}
	\item Exclusion criteria:\\[-0.2cm]
		\begin{itemize}
			\item Inaccessible literature;\\[-0.3cm]
			\item Results that Google Search deems to similar to other results; and\\[-0.3cm]
			\item Vendors tool advertisements.\\[-0.3cm]
		\end{itemize}
\end{itemize}

\begin{figure}[ht]
	\centering
	\includegraphics[width=0.85\textwidth]{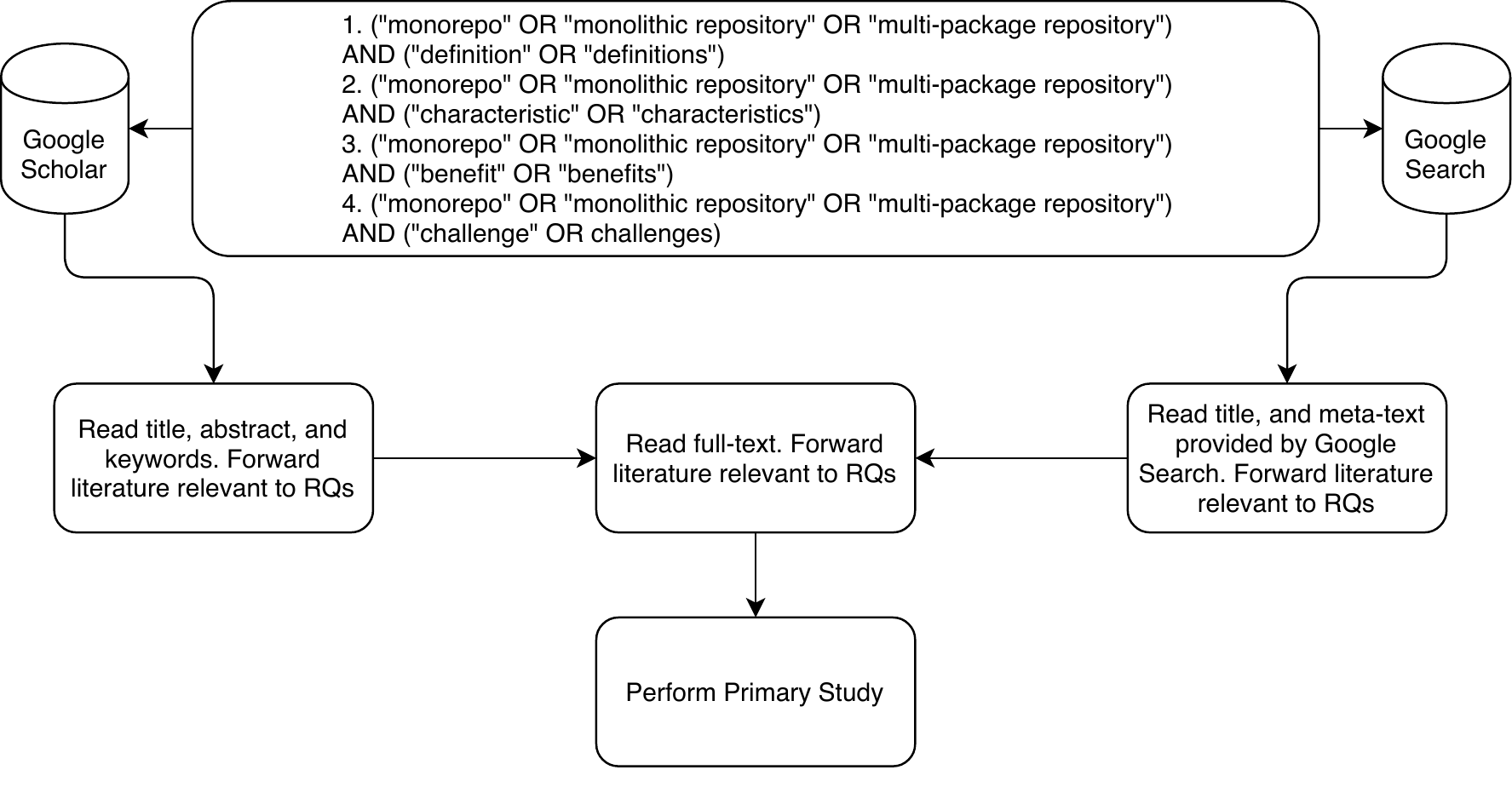}
	\vspace{-10pt}
	\caption{Search process to find relevant literature.}
	\label{fig:search}
\end{figure}

\noindent\textbf{Search Procedure.} As illustrated in Figure~\ref{fig:search}, we first perform an advanced
search in Google Search and Google Scholar. 
For a better focus on each RQs,
we split our search term in four parts.
We analyze 20 pages in total, five pages for each part. 
%
We therefore applied the inclusion and exclusion criteria,
selecting only relevant literature for the primary study. 
In order to identify and incorporate the most relevant grey literature in our MLR, we again use the guidelines defined by Garousi et al.~\cite{Garousi2017GuidelinesFI}.

We performed our search procedure in June, 2018. 
From a total of 255 studies,
we applied the inclusion and exclusion criteria to select 23 studies,
as properly reported in  Table~\ref{tab:primaryStudies}.

\section{Multivocal Literature Review}
\label{results}

Based on the research methodology described in the previous section,
this section reports the results of our multivocal literature review.
%

%


%
\subsection{Definition of Monorepo (RQ~\#1)}

The most common definition of Monorepo is a single repository that contains multiple projects (studies \#4, \#11, \#13, and \#19). These projects can be related or unrelated, but the fact is that they should share the same dependencies. Particularly, studies \#11 and \#13 define monorepos as a single repository that contains more than one logical project. The projects managed in a Monorepo can depend on each other (such as React and the react-dom package) or they can be completely unrelated (such as the Google search algorithm and Angular).

Monorepos can also be classified as \emph{Monstrous monorepos} or \emph{Project monorepos}, according to study \#13. {Monstrous monorepos} regards the sheer size to which monorepos at organizations can grow and {Project monorepos} describes single repositories that are used to manage the core functionality of a project and all of its components. {Google} and {Facebook} repositories are examples of {Monstruous monorepos}, as discussed in studies~\#12 and \#22. {Project monorepos} are commonly adopted by open-source projects with many modules, such as {Babel} and {Ember}. According to study~\#14, the monorepo model allows the maintenance of multiple related packages within a single repository.

\begin{table}[!ht]
	\footnotesize
	\centering
	\caption[]{List of selected primary studies\tablefootnote{We have double checked all URLs on July 2nd, 2018.}}
	\label{tab:primaryStudies}
	\begin{tabular}{@{}c@{\hspace*{4pt}}r@{\hspace*{7pt}}p{13.1cm}@{}}
		\toprule
	Literature                                                                     & \# & Studies                                                                                                                                 \\ \midrule		
& 1                      & \textcolor{mygray}{Anderson, B. (2017). \titulo{Code Repositories and Yak Shaving}. \url{http://iamtherealbill.com/2017/01/repo-yak-shaving/}}\\[0.0475cm]

		& 2                      & \textcolor{mygray}{Belagatti, P. (2016). \titulo{Microservices: Mono repo vs. multiple repositories}. \url{https://jaxenter.com/microservices-mono-repo-vs-multiple-repositories-130148.html}}\\[0.0475cm]

		& 3                      & \textcolor{mygray}{Karanth, D. (2016). \titulo{Microservices: Pros and Cons of Mono Repos}. \url{https://dzone.com/articles/microservices-pros-and-cons-of-mono-repos}}                                                                                  \\[0.0475cm]
		& 4                      & \textcolor{mygray}{Eberlei, B. (2015). \titulo{Monorepos}. \url{https://qafoo.com/talks/15_10_symfony_live_berlin_monorepos.pdf}}                                                                                                                                                              \\[0.0475cm]
		& 5                      & \textcolor{mygray}{Long, C. (2017). \titulo{Multirepo vs Monorepo}. \url{https://chengl.com/multirepo-vs-monorepo/}}                                                                                                                                                                                \\[0.0475cm]
		& 6                      & \textcolor{mygray}{Libbey, B. (2017). \titulo{Monorepo, Manyrepo, Metarepo}. \url{http://notes.burke.libbey.me/metarepo/}}                                                                                                                                                                          \\[0.0475cm]
		& 7                      & \textcolor{mygray}{Johnson, N. (2017). \titulo{Monorepo}. \url{https://www.yonson.io/post/monorepo/}}                                                                                                                                                                                               \\[0.0475cm]
		& 8                      & \textcolor{mygray}{Farina, M. (2016). \titulo{Dangers of Monorepo Projects}. \url{https://dzone.com/articles/dangers-of-monorepo-projects}}                                                                                                                                                         \\[0.0475cm]
		& 9                      & \textcolor{mygray}{Das, S. (2017). \titulo{Code repository for micro-services: mono repository or multiple repositories}. \url{https://medium.com/@somakdas/code-repository-for-micro-services-mono-repository-or-multiple-repositories-d9ad6a8f6e0e}} \\[0.0475cm]
		& 10                     & \textcolor{mygray}{Pendleton, B. (2017). \titulo{Big news in the world of source control}.  \url{http://bryanpendleton.blogspot.com/2017/02/big-news-in-world-of-source-control.html}}                                                                    \\[0.0475cm]
		\multirow{3}{*}{\rotatebox{45}{\bf Grey}}                                                                               & 11                     & \textcolor{mygray}{Saase, S. (2015). \titulo{Monorepos in Git}. \url{https://developer.atlassian.com/blog/2015/10/monorepos-in-git/}}                                                                                                                                                               \\[0.0475cm]
		& 12                     & \textcolor{mygray}{Goode, D. (2014). \titulo{Scaling Mercurial at Facebook}. \url{https:// code.facebook.com/posts/218678814984400/scaling-mercurial-at-facebook/}}                                                                                    \\[0.0475cm]
		& 13                     & \textcolor{mygray}{Oberlehner, M. (2017). \titulo{Monorepos in the Wild}. \url{https://medium.com/@maoberlehner/monorepos-in-the-wild-33c6eb246cb9}}                                                                                                    \\[0.0475cm]
		& 14                     & \textcolor{mygray}{Vepsäläinen, J. (2017). \titulo{Managing Packages Using a Monorepo}. \url{https://survivejs.com/maintenance/ appendices/monorepos/}}                                                                                                   \\[0.0475cm]
		& 15                     & \textcolor{mygray}{Seibel, P. (2017). \titulo{Repo style wars: mono vs multi}. \url{http://www.gigamonkeys.com/mono-vs-multi/}}                                                                                                                                                                     \\[0.0475cm]
		& 16                     & \textcolor{mygray}{Luu, D. (2015). \titulo{Advantages of monorepos}. \url{https://danluu.com/monorepo/}}                                                                                                                                                                                            \\[0.0475cm]
		& 17                     & \textcolor{mygray}{MacIver, D. (2016). \titulo{Why you should use a single repository for all your company’s projects}. \url{https://www.drmaciver.com/2016/10/why-you-should-use-a-single-repository-for-all-your-companys-projects/}}                   \\[0.0475cm]
		& 18                     & \textcolor{mygray}{Fabulich, D. (2017).\titulo{ We’ll Never Know Whether monorepos Are Better}. \url{https://redfin.engineering/well-never-know-whether-monorepos-are-better-2c08ab9324c0}}                                                               \\[0.0475cm]
		& 19                     & \textcolor{mygray}{Szorc, G. (2014). \titulo{On Monolithic Repositories}. \url{https://gregoryszorc.com/blog/2014/09/09/on-monolithic-repositories/}}                                                                                                    \\[0.0475cm]
		& 20                     & \textcolor{mygray}{Beigui, P. (2016). \titulo{Mono-Repos @ Google. Are they worth it?}. \url{https://medium.com/@pejvan/monorepos-85e608d43b57}}                                                                                                         \\[0.0475cm]
		& 21                     & \textcolor{mygray}{Lucido, A. (2017). \titulo{The Journey To Android Monorepo: The History Of Uber Engineering’s Android Codebase Organization}. \url{https://eng.uber.com/android-monorepo/}}                                                           \\[0.0475cm] \midrule
		\multirow{2}{*}[-1.3em]{\rotatebox{45}{\bf Academic}} & 22                     & \textcolor{mygray}{Potvin, R., \& Levenberg, J. (2016). \titulo{Why Google stores billions of lines of code in a single repository}. Communications of the ACM, 59(7), 78-87.}                                                          \\[0.0475cm]
		& 23                     & \textcolor{mygray}{Jaspan, C. et al. (2018). \titulo{Advantages and Disadvantages of a Monolithic Repository - A case study at Google}. 40th International Conference on Software Engineering (ICSE), Software Engineering in Practice (SEIP) Track, 225-234.}                                                    \\[0.0475cm] \bottomrule
	\end{tabular}
	\vspace*{-10pt}
\end{table}

\subsection{Characteristics of Monorepos (RQ~\#2)}

According to studies \#22 and \#23, the five main characteristics of monorepos are:

%
%

\begin{enumerate}[~~~~-]
\setlength{\itemsep}{3pt}
\item \textit{Centralization}: Codebase is in a single
repository encompassing multiple projects.
\item \textit{Standardization}: A shared set of tools govern how
engineers interact with the code, including building,
testing, browsing, and reviewing code.
\item \textit{Visibility}: Code is viewable and searchable by all
engineers in the organization.
\item \textit{Synchronization}: The development process is trunk-based; engineers should always commit to the head of the repository.
\item \textit{Completeness}: Any project in the repository can be built
only from dependencies also checked into the repository.
Dependencies are unversioned; projects must use
whatever version of their dependency at the repository
head.
\end{enumerate}

\subsection{Benefits of monorepos (RQ~\#3)}

This section provides an overview on the benefits of monorepo model. 

\begin{enumerate}[~~~~-]
\setlength{\itemsep}{3pt}
\item \textit{Simplified dependencies}: 
As discussed in studies~\#6 and \#16,
monorepo model proposes to have one universal version number for all projects. Since atomic cross-project commits are possible, the repository is always in a consistent state.  Library versioning is de-emphasized. Instead, a library is expected to maintain a stable API and migrate its callers when this API changes. This depends on being able to make atomic commits. 
\item \textit{Cross-project changes}: Changing APIs that are used in multiple internal projects is more simply in monorepos than in multiple repositories. According to studies~\#16 and \#17, developers can change an API and all its callers in a single commit.
\item \textit{Easy refactoring}: According to studies  \#2, \#16 and \#17, a well-organized unique repository is likely to have modular code and hence refactoring is likely to be easier in monorepos than in multiple repositories. Restructuring is also easier as everything is neatly in one place and easier to understand. 
\item \textit{Simplified organization}: In monorepos, projects can be organized and grouped to be more logically consistent, as described in studies  \#2 and \#16.
\item \textit{Improved overall work culture}: Monorepos encourage the team unification and hence each member can contribute more specifically towards the goals and objectives of the organization, as discussed in study \#2.
\item \textit{Better coordination between developers}: Developers run the entire project on their machine, which helps them understand all services and how they work together. As a result, developers tend to find more bugs locally, before sending pull requests, according to study \#2.
\item \textit{Tooling}: In monorepos, all code 
has a fixed path in a single shared hierarchy, which facilitate building tools that operate on multiple projects, as discussed in studies \#2, \#6, and \#16.
\end{enumerate}

%
%
%

\subsection{Challenges of monorepos (RQ~\#4)}

This section discuss some challenges and trade-offs of monorepos.

\begin{enumerate}[~~~~-]
\setlength{\itemsep}{4pt}
\item \textit{Code health}: In monorepos, according to study~\#22, it is easier to add dependencies. However, (i)~this reduces the incentive for software developers to produce stable and well-defined APIs; (ii)~code cleanup is even more error-prone because
it is common for teams to do not think about their dependency graph; and
(iii)~purposeless dependencies increase project exposure to downstream build breakages, leading to binary size bloating, 
and creating additional work in building and testing.
\item \textit{Codebase complexity}: 
According study \#14, the main challenge of monorepos is to manage all projects in a single repository. 
Although the understanding organization of the code in monolithic repositories is easy,
it is a complex task to determine where new code should be placed.
Besides, study \#7 criticizes monorepos since their high codebase complexity does not necessarily increase productivity.
\item \textit{Tooling investments}: A huge repository requires managing tools to scale. Study~\#1 discusses the high cost of running these tools. 
%


%

\item \textit{Loss of version information}: 
Study \#8 argues that it is dangerous to lose version information. Basically, 
since details of imported libraries can be lost, it may be hard to deal with updates. 

\item \textit{Build, Test Bloat, and Deploy}: Due to its size, studies \#3 and \#8 question the cost of building and testing on monorepos, whereas study \#7 questions the cost of~deploy.

\item \textit{Migration}: Study \#10 introduces \aspas{the monorepo problem}.
It refers to the fact that migrating of many repositories to only one has a high cost, because it is necessary to modularize all code.
This is too critical that study \#21 reports a migration study case.

\end{enumerate}

\section{Threats to Validity}
\label{threatsValidity}
%

%
%
To answer the RQs, we investigated different aspects from monorepos to support
generalization of our discussions. 
This research is mostly based on grey literature, which means that most of the material have not
been subject to rigorous peer-review, as academic research usually is. However, (i)~the inclusion of the grey literature in our review overcame the scarse works available in the digital databases of scientific literature
and (ii)~we analyzed the grey literature in a systematic way by following the guidelines proposed by Garousi~et al.~\cite{Garousi2017GuidelinesFI}.

\section{Conclusion}
\label{conclusion}

This study presented a Multivocal Literature Review on monorepos based mostly on grey literature. We investigate (i)~how monorepos are defined; (ii)~what are the characteristics of monorepos; (iii)~what are the benefits to adopt monorepos; and (iv)~what are the challenges to adopt monorepos.


Regarding~(i), monorepos are usually defined as a single repository that contains multiple related or unrelated projects. In this sense there are two kinds of monorepos: {Monstrous monorepos}, 
which contains several unrelated projects and millions of lines of code, and
and {Projects monorepos}, which contains related components of a specific project.

Regarding~(ii), we can enumerate {\em centralization} since a single repository encompasses multiple projects, {\em visibility} since everything is visible by all contributors, {\em synchronization} since the development process is trunk-based, {\em completeness} since any project can be built using resources available in the repository, and {\em standardization} since engineers usually share the same set of tools in monorepos.

%
%
%
Regarding~(iii), the main benefits include {\em simplified dependencies} since library versioning is easy, {\em simplified organization} since projects are organized in a more consistent way, {\em easy refactoring} since modular repositories foster modular code, {\em improved overall work culture} since monorepos encourage the team unification, {\em   tooling} since a single shared hierarchy facilitate building tools that operate on multiple projects,
{\em better coordination between developers} since developers can easily understand all projects and how they
work together, and {\em better cross-project changes} since an API and its callers can be refactored in a single commit.
%

Regarding~(iv), the main challenges include {\em codebase complexity} since managing all
projects in a single repository is more
difficult, {\em tooling investments} since the cost of running managing tools is large, {\em code health} since unnecessary dependencies can create additional building and testing work, {\em loss of version information} since it is dangerous to
lose some version information, {\em build, test, and deploy} since these tasks can take a long time,  and {\em migration} since the cost of migration of many repositories to only one is high.

%





Several studies compare monorepos with multirepos~(\#2, \#5, \#6, \#9, \#15, and~\#18). These studies argue that choosing monorepos or multirepo is hard because each model has its own set of principles and practices and its own challenges. Monorepos and multirepos not only have different tooling requirements, but also vary in their engineering culture and philosophy. On the one hand, some companies, such as \emph{Netfix}, value freedom and responsability~\cite{netflixCulture}, thus they prefer multirepos. On the other hand, \emph{Google} values consistency and code quality, thus it prefers monorepos.

Ideas for future work include: (i)~to conduct surveys with practitioners to expand our understanding on monorepos;
(ii)~to conduct a study comparing monorepos and multiple repositories models; and
(iii)~to investigate the adoption of monorepos in open-source projects.


\urlstyle{same}
\begin{spacing}{0.87}
\bibliographystyle{plain}
\bibliography{sbc-template}
\end{spacing}

\end{document}